\documentclass[runningheads]{llncs}
\usepackage[utf8]{inputenc} 
\usepackage[T1]{fontenc}    
\usepackage{hyperref}       
\usepackage{url}            
\usepackage{booktabs}       
\usepackage{amsfonts}       
\usepackage{nicefrac}       
\usepackage{microtype}      
\usepackage{xcolor}         
\usepackage{tabularx}
\usepackage{notes-alt} 
\usepackage{multirow}
\usepackage{amsmath}
\usepackage{graphicx}
\graphicspath{ {imgs/} }
\usepackage{svg}
\usepackage{enumitem}
\usepackage{subcaption}
\usepackage{xcolor}
\usepackage[sort,compress]{cite}

\definecolor{blue}{RGB}{17,220,247}
\definecolor{purple}{RGB}{163,115,250}

\usepackage{wrapfig}
\newcommand{\mpara}[1]{\medskip\noindent{\bf #1}}

\newcommand{\ltx}{\textsc{L2x}}
\newcommand{\tabnet}{\textsc{TabNet}}
\newcommand{\cae}{\textsc{Cae}}
\newcommand{\cvar}{\textsc{G-L2x}}
\newcommand{\inv}{\textsc{Invase}}
\newcommand{\lasso}{\textsc{LassoNet}}
\newcommand{\ifg}{\textsc{Ifg}}
\newcommand{\gbdt}{\textsc{Gbdt}}
\newcommand{\dnn}{\textsc{Dnn}}
\newcommand{\selector}{$\zeta$}

\author{Lijun~Lyu\inst{1}\orcidID{0000-0002-7268-4902} 
\and  Nirmal~Roy\inst{1}\orcidID{0000-0003-0860-5269} 
\and Harrie~Oosterhuis\inst{2}\orcidID{0000-0002-0458-9233} \and
Avishek~Anand\inst{1}\orcidID{0000-0002-0163-0739}
}

\institute{
Delft University of Technology, Delft, The Netherlands\\
\email{\{L.Lyu, N.Roy, Avishek.Anand\}@tudelft.nl}
\and Radboud University, Nijmegen, The Netherlands \\
\email{harrie.oosterhuis@ru.nl}
}

\begin{document}

\title{Is Interpretable Machine Learning Effective at Feature Selection for Neural Learning-to-Rank?}
\titlerunning{Is Interpretable ML Effective at Feature Selection for Neural LTR?}

\maketitle

\begin{abstract}

Neural ranking models have become increasingly popular for real-world search and recommendation systems in recent years.
Unlike their tree-based counterparts, neural models are much less interpretable. 
That is, it is very difficult to understand their inner workings and answer questions like \textit{how do they make their ranking decisions?} or \textit{what document features do they find important?}
This is particularly disadvantageous since interpretability is highly important for real-world systems.
In this work, we explore feature selection for neural learning-to-rank (LTR).
In particular, we investigate six widely-used methods from the field of interpretable machine learning (ML) and introduce our own modification, to select the input features that are most important to the ranking behavior. 
To understand whether these methods are useful for practitioners, we further study whether they contribute to efficiency enhancement. 
Our experimental results reveal a large feature redundancy in several LTR benchmarks: the local selection method \tabnet{} can achieve optimal ranking performance with less than 10 features; the global methods, particularly our \cvar{}, require slightly more selected features, but exhibit higher potential in improving efficiency.
We hope that our analysis of these feature selection methods will bring the fields of interpretable ML and LTR closer together.

\end{abstract}

\section{Introduction}
\label{sec:intro}

Learning-to-rank (LTR) is at the core of many information retrieval (IR)
 and recommendation tasks~\cite{liu2009learning}. 
The defining characteristic of LTR, and what differentiates it from other machine learning (ML) areas, is that LTR methods aim to predict the optimal ordering of items.
This means that LTR methods are not trying to estimate the exact relevance of an item, but instead predict relative relevance differences, i.e., whether it is more or less relevant than other items.
Traditionally, the most widely adopted and prevalent LTR methods were based on Gradient Boosted Decision Trees (GBDT)~\cite{burges2010ranknet,wu2010adapting,lightgbm}.
However, in recent years, neural LTR methods have become increasingly popular~\cite{cascade-ranking-cost-aware,pang2020setrank,self-attention-pobrotyn2020context}. 
Recently, \cite{qin2021neural} have shown that neural models can provide ranking performance that is comparable, and sometimes better, than that of state-of-the-art GBDT LTR models on established LTR benchmark datasets~\cite{web30k:DBLP:journals/corr/QinL13,chapelle2011yahoo,dato2016fast}.
It thus seems likely that the prevalence of neural LTR models will only continue to grow in the foreseeable future.

Besides the quality of the results that ranking systems return, there is an increasing interest in building trustworthy systems through interpretability, e.g., by understanding which features contribute the most to ranking results.
Additionally, the speed at which results are provided is also highly important~\cite{bai2019impact,barreda2015unconscious,arapakis2014impact}.
Users expect ranking systems to be highly responsive and previous work indicates that even half-second increases in latency can contribute to a negative user experience~\cite{barreda2015unconscious}.
A large part of ranking latency stems from the retrieval and computation of input features for the ranking model.
Consequently, \emph{feature selection} for ranking systems has been an important topic in the LTR field~\cite{GAS-geng2007feature,gigli2016fast,sun2009rsrank,rahangdale2019deep,pan2009featuretree,xu2014gradient,purpura2021neural}.
These methods reduce the number of features used, thereby helping users understand and greatly reduce latency and infrastructure costs, while maintaining ranking quality as much as possible.
In line with the history of the LTR field, existing work on feature selection has predominantly focused on GBDT and support-vector-machine (SVM) ranking models~\cite{joachims2002optimizing,cascade-ranking-cost-aware}, but has overlooked neural ranking models.
To the best of our knowledge, only two existing works have looked at feature selection for neural LTR~\cite{purpura2021neural,rahangdale2019deep}.
This scarcity is in stark contrast with the importance of feature selection and the increasing prevalence of neural models in LTR.

Outside of the LTR field, feature selection for neural models has received much more attention, for the sake of efficiency~\cite{lemhadri2021lassonet,leonhardt2021learnt}, and also to better understand the model behaviours~\cite{sparcassist,arik2021tabnet}.
Those methods mainly come from the \emph{interpretable} ML field~\cite{du2019techniques,molnar2020interpretable}, where the idea is that the so-called \emph{concrete} feature selection can give insights into what input information a ML model uses to make decisions.
This tactic has already been successfully applied to natural language processing~\cite{zhang:2021:wsdm:expred}, computer vision~\cite{balin2019concrete}, and tabular data~\cite{yoon2018invase,arik2021tabnet}.
Accordingly, there is a potential for these methods to also enable \textit{embedded feature selection} for neural LTR models, where the selection and prediction are optimized simultaneously.
However, the effectiveness of these interpretable ML methods for LTR tasks is currently unexplored, and thus, it remains unclear whether their application can translate into useful insights for LTR practitioners.

The goal of this work is to investigate whether six prevalent feature selection methods -- each representing one of the main branches of interpretable ML field -- can be applied effectively to neural LTR.
In addition, we also propose a novel method with minor modifications. 
Our aim is to bridge the gap between the two fields by translating the important concepts of the interpretable ML field to the LTR setting, and by demonstrating how interpretable ML methods can be adapted for the LTR task.
Moreover, our experiments consider whether these methods can bring efficiency into the practical application by reducing irrelevant input features for neural ranking models.

Our results reveal a large feature redundancy in LTR benchmark datasets, but this redundancy can be understood differently for interpretability and for efficiency:
For understanding the model, feature selection can vary per document and less than 10 features are required to approximate optimal ranking behavior.
In contrast, for practical efficiency purposes, the selection should be static, and then 30\% of features are needed.
We conclude that -- when adapted for the LTR task -- not all, but a few interpretable ML methods lead to effective and practical feature selection for neural LTR.

To the best of our knowledge, this is the first work that extensively studies embedded feature selection for neural LTR.
We hope our contributions bring more attention to the potential of interpretable ML for IR field.
To stimulate future work and enable reproducibility, we have made our implementation publicly available at:
\texttt{\url{https://github.com/GarfieldLyu/NeuralFeatureSelectionLTR}} (MIT license).

\section{Related Work}
\label{related}

\mpara{Learning-to-Rank (LTR)}. 
Traditional LTR algorithms mainly rely on ML models, such as SVMs and decision trees to learn the correlation between numerical input features and human-annotated relevance labels~\cite{li2007mcrank,chapelle2010efficient,freund2003efficient,yue2007support,xu2007adarank,wu2010adapting,lai2012sparse,RankSVM-joachims2006training}. Neural approaches~\cite{burges2005learning,burges2006learning,cao2007learning,xia2008listwise,rigutini2008sortnet,taylor2008softrank} have also been proposed, but did not show significant improvements over traditional non-neural models.
Inspired by the transformer architecture~\cite{vaswani2017attention},  recent works have also adapted self-attention~\cite{pang2020setrank,self-attention-pobrotyn2020context,qin2021neural} and produced the neural LTR methods that outperform LambdaMART~\cite{wu2010adapting}, albeit with a relatively small difference.
It shows that neural rankers can provide competitive performance, consequently, the interest and effort towards neural models for LTR are expected to increase considerably in the near future.

Efficiency is crucial in real-world systems since users expect them to be highly responsive~\cite{bai2019impact,barreda2015unconscious,arapakis2014impact}.
Aside from model execution, the latency of ranking system is largely due to feature construction, as it happens \emph{on-the-fly} for incoming queries.
Thus, efficiency is often reached by reducing (expensive) features.
Previous works~\cite{cascade-ranking-cost-aware,cascade-ranking-early} apply a cascading setup to reduce the usage of expensive features.
Another growing trend in LTR is to design \textit{interpretable models}. Existing methods rely on specific architecture design, such as general additive model (GAM)~\cite{zhuang:2021:wsdm:gam} or a feature-interaction constrained and depth-reduced tree model~\cite{lucchese:2022:sigir:ilmart}.

\mpara{Feature Selection for LTR}.
Feature selection can achieve both efficiency and interpretability~\cite{arik2021tabnet,lemhadri2021lassonet,leonhardt2021learnt,sparcassist}.
By selecting a subset of input features, the input complexity of models is reduced while maintaining competitive performance.
This helps with (1) efficiency as it avoids unnecessary construction of features~\cite{lemhadri2021lassonet,leonhardt2021learnt}, and (2) interpretability as fewer input features are involved in prediction~\cite{sparcassist,arik2021tabnet}.

Existing feature selection methods in LTR are classified commonly as \textit{filter}, \textit{wrapper} and \textit{embedded} methods~\cite{GAS-geng2007feature,gigli2016fast,purpura2021neural}.
Filter and wrapper methods are applied to given static ranking models which are not updated in any way;  filter methods are model-agnostic~\cite{GAS-geng2007feature} while wrapper methods are designed for a particular type of model~\cite{gigli2016fast}.
In this work we will focus on the third category, embedded methods, where feature selection is performed simultaneously with model optimization.
Most embedded methods are limited to particular model designs such as SVMs~\cite{lai2012fenchelrank,lai2013fsmrank,laporte2013nonconvex} or decision trees~\cite{pan2009featuretree,xu2014gradient,ilmart}.
To the best of our knowledge, only two methods are designed for neural LTR~\cite{rahangdale2019deep,purpura2021neural}: one applies group regularization methods~\cite{rahangdale2019deep} to reduce both input and other model parameters; the other~\cite{purpura2021neural} uses the gradients of a static ranking model to infer feature importance, and thus it belongs to the \emph{filter} category. We do not investigate these two methods further,
as the focus of this work is on \emph{embedded} input feature selection methods.

\mpara{Interpretable Machine Learning}.
 The earliest work in interpretable ML attempted to explain a trained model in \emph{post-hoc} manner, mainly relying on input perturbations~\cite{ribeiro2016should,lundberg2017unified}, gradients~\cite{simonyan2013deep,integratedgradients} and so on~\cite{shrikumar2017learning}.
 In parallel, more recent works advocated intrinsically interpretable models, that are categorized as \emph{interpretable-by-design} methods~\cite{rudin:stopexplain,selfexplaining,gam-cognitive}.
 For neural networks, explaining the decision path is challenging due to the large set of parameters.
Therefore, the more prevalent choice for intrinsic interpretable neural models is to shift the transparency to the input features. Namely, the final prediction comes from a subset selection of input elements, e.g., words or pixels and the rest irrelevant features are masked out~\cite{zhang:2021:wsdm:expred,chen2018learning,leonhardt2021learnt}.
Importantly, this selection decision can be learned jointly with the predictive accuracy of a model. Thereby, we limit our research focus in intrinsic interpretable ML models.

Due to the discrete nature of selection, many approaches such as L2X~\cite{chen2018learning}, \textit{Concrete AutoEncoders} (CAE)~\cite{balin2019concrete}, \textit{Instance-wise Feature grouping} (IFG)~\cite{masoomi2020instance} apply Gumbel-Softmax sampling~\cite{gumbel-jang-2016} to enable backpropagation through the feature selection process.
Alternatively, regularization is also a commonly-used feature selection approach in traditional ML algorithms~\cite{lasso-tibshirani1996regression,lai2012sparse}, and is applicable to neural models, i.e., with INVASE~\cite{yoon2018invase} or LassoNet~\cite{lemhadri2021lassonet}.
Moreover, TabNet~\cite{arik2021tabnet} applies both regularization and the sparsemax activation function~\cite{martins2016softmax} to realize sparse selection.
These approaches have been successfully applied in language, vision and tabular domains, and suggested that the resulting feature selections substantially improved the user understanding of models and datasets~\cite{interpretation-usage,userstudy-survey}.

Despite their success in other domains, we find that the above-mentioned feature selection methods for neural models (L2X, CAE, IFG, INVASE, LassoNet and TabNet) have not been studied in the LTR setting.
In response, we hope to bridge this gap between the interpretable ML and the LTR field by adapting and applying these methods to neural ranking models.

\section{Background}
\label{sec:prelims}
\subsection{Learning-to-Rank (LTR)}
\label{sec:prelims-ltr}

The LTR task can be formulated as optimizing a scoring function $f$ that given item features $x$ predicts an item score $f(x) \in \mathbb{R}$, so that ordering items according to their scores corresponds to the optimal ranking~\cite{liu2009learning}.
Generally, there are relevance labels $y$ available for each item, often these are labels provided by experts where $y\in\{0,1,2,3,4\}$~\cite{chapelle2011yahoo,mq2008:qin2010letor,web30k:DBLP:journals/corr/QinL13}.
Given a training set $\mathcal{D}_q = \{(x_i, y_i)\}^{N_q}_{i=1}$ for a single query $q$, optimization is done by minimizing a LTR loss, for instance, the \emph{softmax cross entropy loss}~\cite{bruch2019analysis,cao2007learning}:
\begin{equation}
    \mathcal{L}(f \mid \mathcal{D}_q) = -\frac{1}{|\mathcal{D}_q|} \underset{(x, y)\in \mathcal{D}_q}{\sum} \sum_{i=1}^{N_q} y_i \log \sigma (x_i \mid f, \mathcal{D}_q),
\label{eq:LTR}
\end{equation}
where $\sigma$ is the softmax activation function:
\begin{equation}
\sigma (x \mid f, \mathcal{D}_q) = \frac{\exp(f(x))}{\sum_{x' \in \mathcal{D}_q} \exp(f(x'))}.
\end{equation}
The resulting $f$ is then commonly evaluated with a ranking metric, for instance, the widely-used normalized discounted cumulative gain metric (NDCG)~\cite{jarvelin2002cumulated}.

\subsection{Properties of Feature Selection Methods}
\label{sec:lts}
\begin{table}[t]
    \centering
   \caption{Properties of feature selection methods from the interpretable ML field as discussed in Section~\ref{sec:prelims}.}
    \label{tab:methods_summary}
    \begin{tabular*}{\textwidth}{lcccccc}
    \toprule
      
        \multicolumn{1}{l |}{\bf Method}   &  \multicolumn{1}{c |}{ Global} &  \multicolumn{1}{c |}{ Local} & \multicolumn{1}{c |}{ Sampling}&  \multicolumn{1}{c |}{ Regularization} &  \multicolumn{1}{c |}{ Fixed-Budget} & \multicolumn{1}{c}{ Composable} \\
        \midrule

        \ltx{}  \cite{chen2018learning} & & \checkmark & \checkmark & & \checkmark & \checkmark\\

        \inv{} \cite{yoon2018invase} & &\checkmark & \checkmark & \checkmark & &\checkmark \\

        \cae{}  \cite{balin2019concrete}  & \checkmark & & \checkmark & &\checkmark  & \checkmark\\

        \ifg{} \cite{masoomi2020instance}  & & \checkmark & \checkmark & & &\checkmark \\

        \lasso{} \cite{lemhadri2021lassonet}  & \checkmark & & & \checkmark & &\checkmark\\

        \tabnet{} \cite{arik2021tabnet} & &\checkmark & & \checkmark& & \\

        \cvar{} (ours) & \checkmark & & \checkmark & &\checkmark  & \checkmark\\
    
       \bottomrule
        
    \end{tabular*}
\end{table}

As discussed before, feature selection is used in the interpretable ML field to better understand which input features ML models use to make their predictions.
Furthermore, feature selection is also important to LTR for increasing the efficiency of ranking systems.
However, selecting a subset of input features without compromising the model performance is an NP-hard problem, since the number of possible subsets grows exponentially with the number of available features~\cite{gigli2016fast,purpura2021neural}.
As a solution, the interpretable ML field has proposed several methods that approach feature selection as an optimization problem.
We will now lay out several important properties that can be used to categorize these methods, which will be elaborated in next section.

\mpara{Global vs.\ local.} 
Global methods select a single subset of features for the entire dataset, whereas local methods can vary their selection over different items.

\mpara{Composable vs.\ non-composable.} 
Non-composable methods are designed for a specific model architecture, and therefore, they can only perform feature selection for those models.
Conversely, composable methods are not constrained to specific architectures, and thus, they work for any (differentiable) model.

\mpara{Fixed-budget vs.\ budget-agnostic.} 
Fixed-budget methods work with a pre-defined selection budget, i.e., what number of features should be selected in total, or a cost per feature and a maximum total cost for the selection.
Their counterparts are budget-agnostic methods that do not use an explicit budget, consequently, one has to carefully fine-tune their hyper-parameters to achieve a desired performance-sparsity trade-off.

\mpara{Sampling-based vs.\ regularization-based.}
As their names imply, sampling-based methods optimize a sampling procedure to perform the feature selection, whereas regularization-based methods use an added regularization loss to stimulate sparsity in feature selection.
While these groups apply very different approaches, whether one is significantly more useful for LTR purposes than the other remains unknown.

\section{Feature Selection from Interpretable ML for LTR}
\label{sec:methods}

\begin{figure*}[t]
    \centering
       \includegraphics[width=0.95\linewidth]{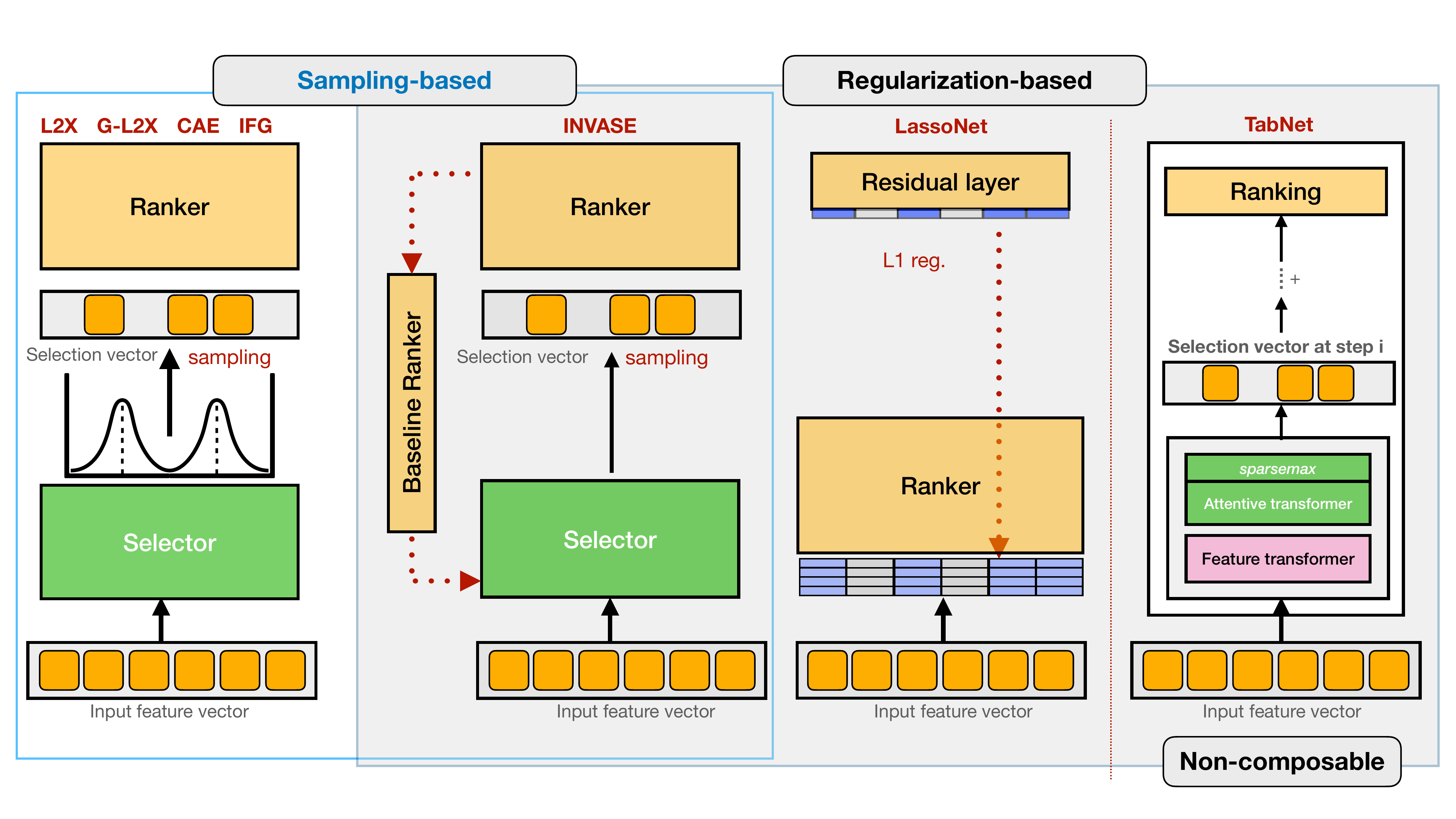}
     \caption{Methods overview, as described in Section~\ref{sec:methods}.}
    \label{figs:lts-ltr}
    \vspace{-0.8\baselineskip}
\end{figure*}

In this section, we present a brief technical overview of our selection of six interpretable ML methods and their adaption to neural LTR models, and propose our \cvar{} method based on a minor modification.
While any ranking loss can be chosen, we use a listwise softmax cross entropy (Eq.~\ref{eq:LTR}) with all methods, for the sake of simplicity. Therefore, the training of each query is conducted after generating the output of all documents associated with the query.
Table~\ref{tab:methods_summary} highlights the properties of all methods and
Figure~\ref{figs:lts-ltr} provides a visual overview to accompany this section.

\subsection{Sampling-based Feature Selection}
\label{sec:sampling-based}

Sampling-based approaches use a two-stage architecture consisting of a \emph{selector} that generates a sparse selection over the input features; and a \emph{ranker} that only takes selected features as its input, in the form of a masked vector $\hat{x}$.

The training of a ranker follows conventional LTR, i.e., Eq~\ref{eq:LTR} with $x_i$ replaced by $\hat{x_i}$.
But the optimization of a selector (\selector{}) is not as straightforward;
Usually, \selector{} constructs a probability distribution $\mathbf{p} = [p_1, p_2, \cdots p_{d}]$, indicating a selection probability per feature. 
However, the ranker uses a concrete selection $m \in \{0,1\}^d$ from the probability distribution, and this concrete operation does not allow optimization of the selector via backpropagation.
The common solution is to generate a differentiable approximation $\tilde m$,
by \emph{concrete relaxation} or the Gumbel-Softmax trick~\cite{gumbel-jang-2016}.
Namely, the selection of $p_i$ can be approximated with the differentiable $c_i$ as:
\begin{equation}
    c_i = \frac{exp\{(\log p_i + g_i)/\tau\} }{ \sum_{j=1}^d exp\{(\log p_j + g_i)/\tau\}},
\label{eq:gumbel}
\end{equation}
where $g$ is the Gumbel noise and $\tau \in \mathbb{R}^{>0}$ is the temperature parameter.
Now, the selector \selector{} can be optimized with stochastic gradient descent by using $\tilde{m}$.
The following four sampling-based methods apply this overall procedure, but differ in how they generate $\mathbf{p}$ and $\tilde m$.

\mpara{Learning to explain (\ltx{}).} 
\ltx{}~\cite{chen2018learning} is a local selection method since its neural selector generates a probability distribution $\mathbf{p}$ for each individual input instance. To generate $\tilde m$, \ltx{} repeats the sampling procedure  $k$ times (Eq.~\ref{eq:gumbel}), and subsequently, uses the maximum $c_i$ out of the $k$ repeats for the $i_{th}$ element in $\tilde m$.
The intention behind this maximization step is to make the top-$k$ important features more likely to have high probability scores (ideally close to 1).

\mpara{Global learning to explain (\cvar{}, ours).}
As a counterpart, we propose a global method \cvar based on \ltx{}.
Our change is straightforward, where \ltx{} generates a different distribution $\mathbf{p}$ for each item, we apply the same $\mathbf{p}$ to all items.
In other words, \cvar{} includes a global selector layer \selector{} ($\zeta \in \mathbb{R}^d$) to simulate $\mathbf{p}$, and sampling is conducted in the same way as \ltx{} on the selector weights.
Thereby, \cvar{} will select the same features for all items in the dataset.

\mpara{Concrete autoencoder (\cae{}).} 
 \cae{}~\cite{balin2019concrete} is a global method where the selector is the encoder part of an auto-encoder model~\cite{kingma2013auto}.
 Specifically, the selector compresses the input into a smaller representation $\hat{x}$, by linearly combining selected features, i.e. $x^{\top}\tilde{m}$, where  $\tilde{m} \in \mathbb{R}^{k\times d}$ can be viewed as approximated k-hot concrete selection, sampled from the selector weights ($\zeta \in \mathbb{R}^{k\times d}$). Therefore, \cae{} might result in repetitive selection, and the input dimension to the predictor is reduced to $k$.

\mpara{Instance-wise Feature Grouping (\ifg{}).} 
\ifg{}~\cite{masoomi2020instance} applies a similar approach as \ltx{}, but clusters features into groups and then selects $k$ feature groups for prediction.
\ifg{} first assigns a group for each feature via Gumbel-sampling, and then makes a feature selection by Gumbel-sampling $k$ out of the resulting groups.
This grouping decision is also guided by how rich the selected features are to recover the original input.
Therefore, apart from the ranking objective, \ifg{} jointly optimizes an additional input restoring objective as well (similar to auto-encoders~\cite{kingma2013auto}).
\ifg{} is agnostic to the number of selected features and the group sizes, it can produce oversized groups and very large selections.

\subsection{Regularization-based Feature Selection}
Instead of the budget-explicit feature selection, regularization-based methods induce sparsity through implicit constraints enforced by regularization terms in the training objective.
We propose modifications to three existing methods to make them applicable to the LTR setting.

\mpara{\inv{}.}
We consider \inv{}~\cite{yoon2018invase} to be a hybrid approach involving both sampling and regularization.
Built on the same structure as \ltx{}, the selector of \inv{} generates a \textit{boolean/hard mask} $m$ (instead of the approximation $\tilde{m}$) via Bernoulli sampling to train the predictor.
Since this disables backpropagation, \inv{} uses a customized loss function that does not need the gradients from the predictor to train the selector. The idea is to apply another individual baseline predictor model that takes the full-feature input, simultaneously with the predictor that takes the masked input. The loss difference between the two predictors is used as a scale to train the selector. Meanwhile, the L1 regularization is applied to the selector output to enforce selection sparsity. Ultimately, \inv{} will push the selector to output a small set of selections which leads to the most similar predictions as using all features.

\mpara{\lasso{}.} As the name suggests, \lasso{}~\cite{lemhadri2021lassonet} adapts a traditional Lasso (L1 regularization)~\cite{lasso-tibshirani1996regression} on the first layer of a neural model to eliminate unnecessary features. The challenge with neural models is, all weights corresponding to a particular feature entry in the layer have to be zero in order to mask out the feature. Towards this, \lasso{} adds a residual layer with one weight per input feature to the original model to perform as the traditional Lasso. Then, after every optimization step, \lasso{} develops a proximal optimization algorithm to adjust the weights of the first layer, so that all absolute elements of each row are smaller than the respective weight of residual layer corresponding to a specific feature. Thereby, \lasso{} performs global selection and the sparsity scale is adjusted by the L1 regularization on the residual layer weights.

\mpara{\tabnet{}.}
Unlike the previous methods,  \tabnet{}~\cite{arik2021tabnet} is non-composable and tied to a specific tree-style neural architecture. It imitates a step-wise selection process before it outputs the final prediction based only on the selected features.
Each step has the same neural component/block but with its own parameters, thus the model complexity and selection budget grow linearly with the number of steps. At each step, the full input is transformed by a feature transformer block first, and then an attentive transformer block conducts feature selection by sparsemax activation~\cite{martins2016softmax}, as the weights in the resulting distribution corresponding to unselected features are zeros. 
The final prediction is aggregated from all steps to simulate ensemble models. A final mask $m$ is a union of selections from all steps, and the entropy of the selection probabilities is used as the sparsity regularization.

\section{Experimental Setup}
\label{sec:experiments}

Since feature selection can be applied in various manners and situations, we structure our experiments around three scenarios:
\begin{itemize}[leftmargin=*,]
    \item \textit{Scenario 1: Simultaneously train and select}.
    Both the ranking model and the feature selection are learned once and jointly. The methods are evaluated by the performance-sparsity trade-off.
    It is the standard setup for evaluating embedded feature selection methods in the interpretable ML field~\cite{rahangdale2019deep,masoomi2020instance}.
    
    \item \textit{Scenario 2: Train then select with an enforced budget}.
    Practitioners generally set hard limits to the computational costs a system may incur and the efficiency of the system can be greatly enhanced if it only requires a much smaller amount of features to reach competitive performance. Following the previous scenario, we evaluate the trained model with test instances where only a fixed amount of features (which the method deems important and selects frequently during training) are presented and the rest are masked out. The resulting ranking performance and the costs of computing the required features indicate how practical the method is in efficiency improvements.

\end{itemize}

\mpara{Datasets and preprocessing}.
We choose three public benchmark datasets: MQ2008 (46 features)~\cite{mq2008:qin2010letor}, Web30k (136 features)~\cite{web30k:DBLP:journals/corr/QinL13} and Yahoo (699 features)~\cite{chapelle2011yahoo}, to cover varying  numbers of available features. 
We apply a $\log_{1p}$ transformation to the features of Web30k, as suggested in~\cite{qin2021neural}.
Yahoo contains cost labels for each feature, for Web30k we use cost estimates suggested by previous work~\cite{cascade-ranking-cost-aware}.\footnote{MQ2008 is omitted from cost analysis since no associated cost information is available.}
All reported results are evaluated on the held-out test set partitions of the datasets.

\mpara{Models}.
We use a standard feed-forward neural network with batch normalization, three fully-connected layers and tanh activation as the ranking model, denoted as \dnn{}.
According to the findings in \cite{qin2021neural}, this simple model performs closely to the most effective transformer-based models, but requires much less resources to run.
The selector models of \ltx{} and \inv{} have the same architecture, and as the only exception, \tabnet{} is applied with its own unalterable model (see Section~\ref{sec:methods}).

\mpara{Implementation}.
Our experimental implementation is done in \textit{PyTorch Lightning}~\cite{falcon2019pytorch}.
For \tabnet{} and \lasso{} existing implementations were used.\footnote{\url{https://github.com/dreamquark-ai/tabnet}; \url{https://github.com/lasso-net/lassonet}}
We created our own implementations for the rest of the methods.
\begin{table*}[t]
\caption{Results of ranking performance and feature sparsity for methods applied in Scenario 1.
For comparison, we also include \gbdt{}~\cite{lightgbm,bruch2021alternative} and \dnn{} baselines without feature selection as upper bound.
\#F denotes the number of selected features.
Reported results are averaged over $5$ random seeds (\textit{std} in parentheses).
Bold font indicates the highest performing selection method; the $\star$ and underlines denote scores that are \emph{not} significantly outperformed by \gbdt{} and the bold-score method, respectively ($p > 0.05$, paired t-tests using Bonferonni's correction).}
\label{tab:best_results} 
\centering
\resizebox{\textwidth}{!}{
\begin{tabular}{lllllllllr}
\toprule
\textbf{Listwise} & \multicolumn{3}{c}{\textbf{MQ2008 NDCG@k} (\%)} & \multicolumn{3}{c}{\textbf{Web30k NDCG@k} (\%)} &
\multicolumn{3}{c}{\textbf{Yahoo NDCG@k} (\%)} \\

\textbf{loss} & @1 &  @10& \#F & @1 & @10& \#F & @1 & @10 & \#F\\
\midrule

 \multicolumn{10}{l}{\bf \textbf{Without feature selection.}} \\

 \gbdt{} & 69.3 (2.5) & 80.8 (1.7) & 46& 50.4 (0.1)  & 52 (0.1) & 136 & 72.2 (0.1) & 79.2 (0.1) & 699\\

 \dnn{}& \underline{66.2}$^{\star}$ (2) & \underline{80.2}$^{\star}$ (0.6) & 46 & \underline{46.1} (0.6) & 47.7 (0.2)  &  136 & 69.4 (0.3)  & 76.9 (0.1)  & 699 \\
\midrule

\multicolumn{9}{l}{\bf \textbf{Fixed-budget feature selection using the \dnn{} ranking model.}} \\
  
\cae{} & \underline{63.0} (1.1) & 78.7 (0.5) & 4 & 32.9 (2.9) & 36.6 (2.2) & 13 & 59.2 (0.2)  & 69.5 (0.1) & 6\\

\cvar{} & \underline{63.8}$^{\star}$ (1.3) & \underline{79.1}  (0.4) & 4 & 41.1 (0.9) & 44.4 (0.3) & 13 & 65.4 (0.1) & 74.0 (0.0) &6 \\

\ltx{} & \underline{63.0} (2.1)  & \underline{78.7} (0.7) & 4 & 34.5 (2.4) & 39.7 (1.9) & 13 & 61.9 (1.1) & 73.2 (0.3) & 6\\

\midrule
\multicolumn{9}{l}{\bf \textbf{Budget-agnostic feature selection using the \dnn{} ranking model.}} \\

\inv{} & \underline{62.5} (2.2) & \underline{77.5} (2.1) & 5 (2) & 15.1 (0.0) & 22.1 (0.0) & 0& 38.7 (0.0) & 57.8 (0.0) & 0\\

\ifg{} & \textbf{66.4}$^{\star}$ (0.9) & \textbf{80.4}$^{\star}$ (0.5) & 20 (2) & 32.5 (5.3) & 37.5 (5.3) & 72 (30) & 69.6 (0.2) & 77.1 (0.2) & 58 (3)\\

\lasso{} & \underline{64.7}$^{\star}$ (2.2) & \underline{79.3} (1.2) & 6 (3) & 39.4 (0.8) &  42.1 (0.3) & 8 (2) & 63.1 (2.3) & 72.4 (1.5) & 12 (4)\\
\midrule
\multicolumn{9}{l}{\bf \textbf{Budget-agnostic feature selection using a method-specific ranking model.}} \\
\tabnet{} & \underline{64.7}$^{\star}$ (2.7) &  \underline{78.2} (1.2) & 7 (3) & \textbf{47.0} (0.4)  & \textbf{49.2} (0.1) & 8 (1) & \textbf{70.2} (0.4)  & \textbf{ 77.7} (0.1) & 6 (1)\\
\bottomrule 
\end{tabular}
}
\end{table*}

\begin{figure*}[ht]
\centering
{\renewcommand{\arraystretch}{0.0}
\setlength{\tabcolsep}{0.05cm}
\begin{tabular}{l l l l }
& 
 \multicolumn{1}{c}{\hspace{0.24cm} \small  MQ2008}
&
 \multicolumn{1}{c}{\hspace{0.13cm} \small Web30k}
&
 \multicolumn{1}{c}{\hspace{0.5cm} \small Yahoo}
 
\\
\multirow{1}{2mm}{\raisebox{4\normalbaselineskip}[0pt][0pt]{\rotatebox[origin=c]{90}{\small NDCG@10}}}&
\includegraphics[scale=0.34]{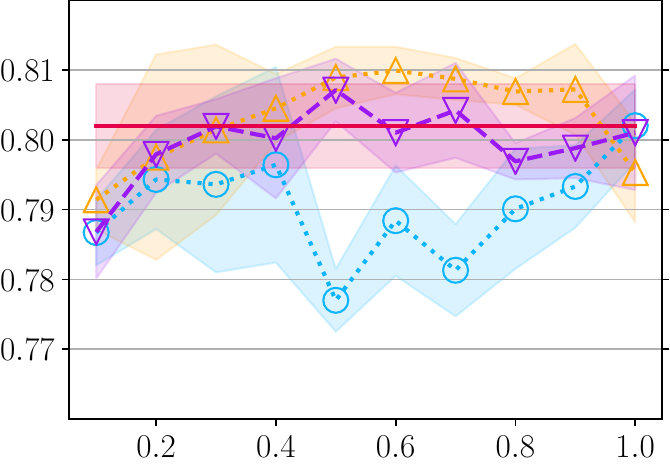} &
\includegraphics[scale=0.34]{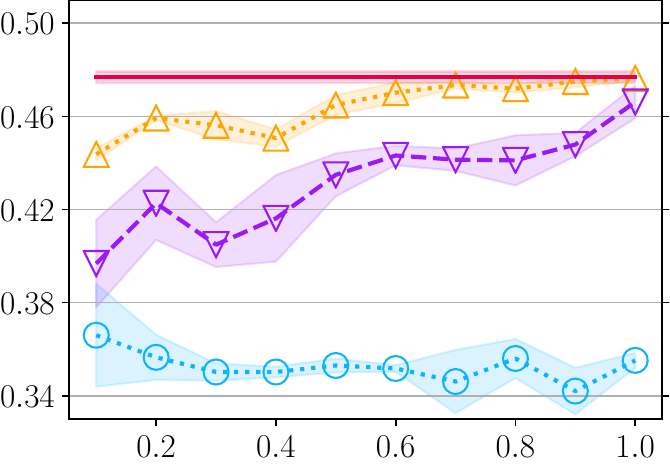} &
\includegraphics[scale=0.34]{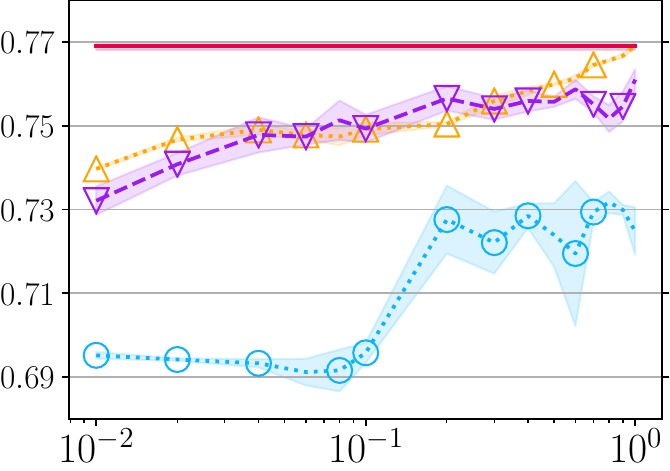}
\\
 \multicolumn{4}{c}{{\includegraphics[scale=0.45]{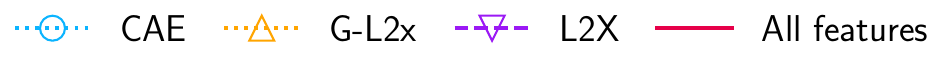}} } \\
%
%
\end{tabular}
}
%
%
\caption{Results of three fixed-budget methods applied to scenario 1. The x-axis indicates the pre-specified percentile of selected features ($k$). The shaded area shows the standard deviation over 5 random seeds.}
\label{figs:controllable}
%
%
\end{figure*}

\section{Results}

We report the findings in this section, aiming to answer two questions: (1) \textit{how effective are investigated methods in the ranking setup?} and (2) \textit{can those methods improve efficiency?}  Each question corresponds to one of the scenarios described in Section~\ref{sec:experiments}.

\mpara{Simultaneous Optimization and Selection}. 
We begin by investigating the effectiveness of the feature selection methods when applied to Scenario 1, where feature selection and model optimization are performed simultaneously.
The results for this scenario are displayed in Table~\ref{tab:best_results} and Figure~\ref{figs:controllable}.

Table~\ref{tab:best_results} shows the ranking performance and the respective feature sparsity of all feature selection methods and two baselines without any selection as the upper-bound reference.
For fixed-budget methods, the budgets were set to 10\% of the total number of features for MQ2008 and Web30k, and 1\% for Yahoo (the results with varying budgets are displayed in Figure~\ref{figs:controllable}).
Since the sparsity of budget-agnostic methods is more difficult to control, we performed an extensive grid search and used the hyper-parameters that produced the highest ranking performance with a comparable feature sparsity as the other methods. 

The results in Table~\ref{tab:best_results} show that not all feature selection methods are equally effective, and their performance can vary greatly  over datasets.
For instance, on MQ2008 all methods perform closely to the baselines, with only a fraction of the features.
However, this is not the case for bigger datasets like Web30k and Yahoo.
In particular, \inv{} selects no features at all due to big uncertainty in selection (for this reason, we omit \inv{} from all further comparisons). On the other hand, \ifg{} performs poorly in inducing sparsity, mainly because of its input reconstruction objective, whereas the ranking performance is not substantially better than the rest of methods. Additionally, \cae{} does not seem effective either, and furthermore, increasing the selected features does not always result in better ranking performance (cf.\ Figure~\ref{figs:controllable}).
This is most-likely because \cae{} samples with replacement, and thus the same features can be selected repeatedly. 

In contrast, the other two sampling-based methods \ltx{} and \cvar{} are designed to avoid the repetitive selection issue.
Overall, the global selection \cvar{} outperforms the local counterpart \ltx{}, possibly because global selection generalizes better to unseen data.
Another global method \lasso{} is also inferior to \cvar{}, mainly due to the difficulties in sparsity weight tuning and manually adjusting weights in the input layer.

Lastly, \tabnet{} shows the best performance-sparsity balance across all datasets, and even outperforms the \dnn{} baseline.
Although, the comparison between \tabnet{} and \dnn{} is not completely fair, as they optimize different neural architectures.
It does reveal large feature redundancies in these datasets:
\tabnet{} uses <10\% of features on Web30k and 1\% on Yahoo, yet still beats the \dnn{} baseline with all features.

To summarize,
we find that the local method \tabnet{} is the most effective at balancing ranking performance and sparsity. Slightly inferior but competitive enough is the global method \cvar{}, which reached $>95\%$ of baseline performance with only 1\% features on Yahoo and $>93\%$ with 10\% on Web30k.

\begin{figure}[t]
\centering
{\renewcommand{\arraystretch}{0.4}
\setlength{\tabcolsep}{0.04cm}
\begin{tabular}{ l l  l l}
\multicolumn{2}{c}{ Feature Cost } & 
\multicolumn{2}{c}{ Ranking Performance (NDCG@10)\vspace{1mm} }
\\
 \multicolumn{1}{c}{ \small Web30k}
&
 \multicolumn{1}{c}{ \small Yahoo}
 & 
 \multicolumn{1}{c}{ \small Web30k}
&
 \multicolumn{1}{c}{ \small Yahoo}
\\
%
\includegraphics[scale=0.31]{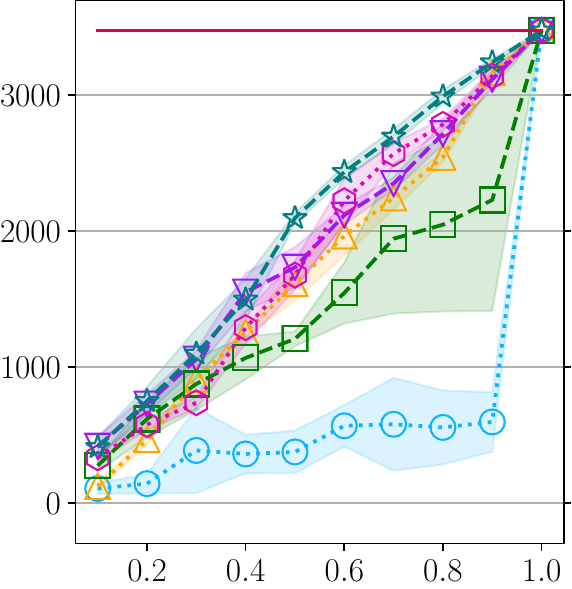} &
\includegraphics[scale=0.31]{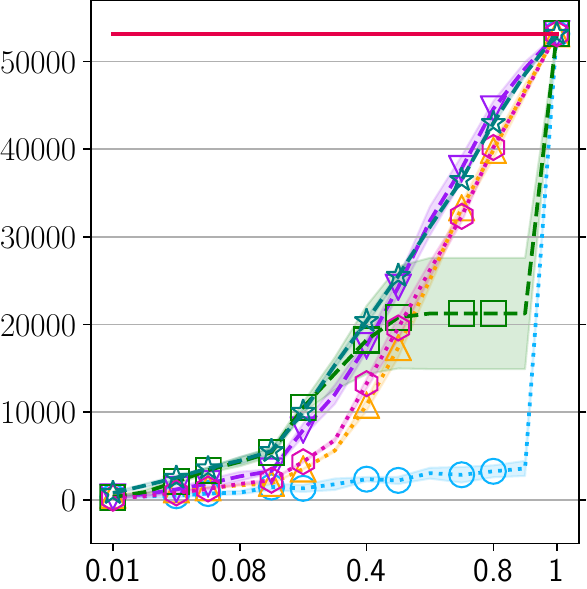} &
\includegraphics[scale=0.31]{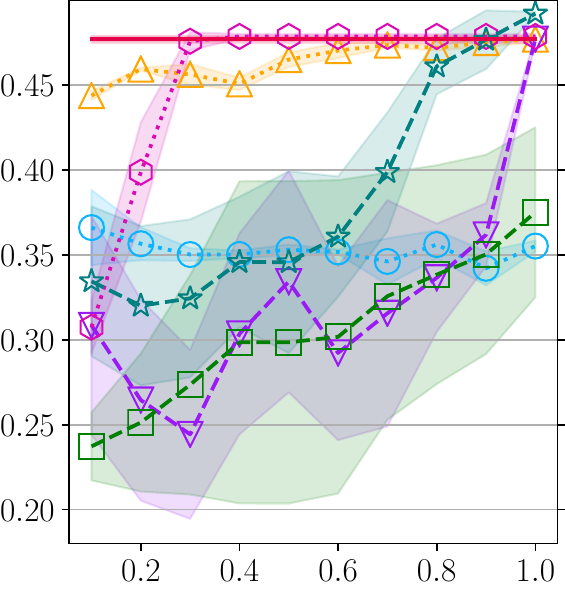} & 
 \includegraphics[scale=0.31]{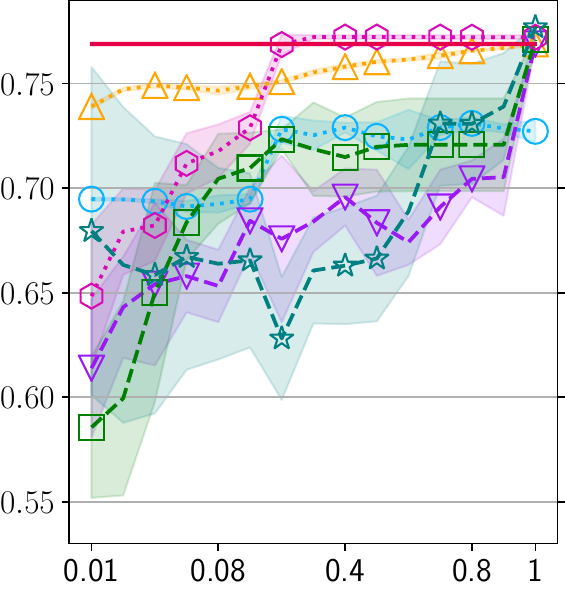}
\\
\multicolumn{4}{c}{
\includegraphics[scale=0.45]{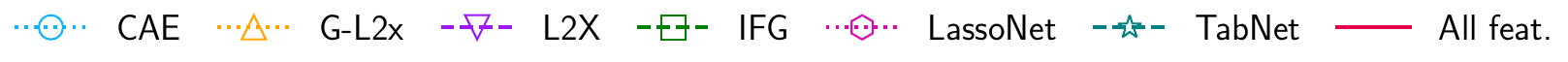}
}
\end{tabular}
}
%
%
\caption{Scenario 2. Feature cost (left two) and ranking performance (right two) under incomplete input. The x-axis indicates how many percentages of features are present in the input, to test the trained ranking model. Note this differs from specifying $k$ during training for fixed-budget methods in scenario 1.
}
\label{figs:ndcg_cost}
\end{figure}

\mpara{Feature Selection for Trained Ranking Models}.
Next, we evaluate the methods in Scenario 2, where only a specified budget (i.e., a given number of features) of features are present in the test input.
Figure~\ref{figs:ndcg_cost} displays both the ranking performance and the total feature cost for varying degrees of sparsity.
The costs represent the time it requires to retrieve the selected feature sets, and allow us to estimate the actual efficiency improvements they provide.

Unlike previous scenario, all local methods including \tabnet{}, are no longer able to maintain superior performance.
This is because for local methods, the selection is made conditioned on full input information,
and an incomplete input could affect the selection and thus disrupt its prediction performance.

In contrast, global methods are immune to input changes. Therefore, \cae{} is still not performing well as it did in Scenario 1; 
 \cvar{} and \lasso{} provide the best overall performance under small costs.
\lasso{} maintains baseline performance with less than 40\% of features on both datasets, while \cvar{} outperforms \lasso{} when selected features are less than 30\%. Meanwhile, it also shows \lasso{} tends to select more costly features than \cvar{}.

To conclude, we find that global methods \cvar{} and \lasso{} perform the best in Scenario 2, where upcoming query inputs are masked under enforced feature budgets. Particularly, \cvar{} is superior in both ranking and computing cost when the feature budget is small. This translates to substantial efficiency improvements in practical terms, as ranking performance is maintained by selected features only.

\section{Conclusion}

The main goal of this work is to bring the interpretable ML and the LTR fields closer together.
To this end, we have studied whether feature selection methods from the interpretable ML are effective for neural LTR, for both interpretability and efficiency purposes.

Inspired by the scarcity of feature selection methods for neural ranking models in previous work,
we adapted six existing methods from the interpretable ML for the neural LTR setting, and also proposed our own \cvar{} approach.
We discussed different properties of these methods and their relevance to the LTR task.
Lastly, we performed extensive experiments to evaluate the methods in terms of their trade-offs between ranking performance and sparsity, in addition, their efficiency improvements through feature cost reductions.
Our results have shown that several methods from interpretable ML are highly effective at embedded feature selection for neural LTR. 
In particular, the local method \tabnet{} can reach the upper bound with less than 10 features; the global methods, in particular \cvar{}, can reduce feature retrieval costs by more than 70\%, while maintaining 96\% of performance relative to a full feature model.

We hope our investigation can bridge the gap between the LTR and interpretable ML fields. The future work can be developing more interpretable and efficient ranking systems, and how that interpretability could support both practitioners and the users of ranking systems.

\subsubsection{Acknowledgements}

This work is partially supported by German Research Foundation (DFG), under the Project IREM with grant No.\,AN 996/1-1, and by the Netherlands Organisation for Scientific Research (NWO) under grant No.\ VI.Veni.222.269.


\bibliographystyle{splncs04}
\bibliography{main} 


\end{document}